\documentclass[useAMS,usenatbib]{mn2e}

\usepackage{times}
\usepackage{graphics,epsfig}
\usepackage{graphicx}
\usepackage{amsmath}
\usepackage{amssymb}

\newcommand{\kms}{km~s$^{-1}$}
\newcommand{\arcs}{$^{\prime\prime}$}
\newcommand{\sgas}{\ensuremath{\Sigma_{\rm gas}}}
\newcommand{\shi}{\ensuremath{\Sigma_{\rm HI}}}
\newcommand{\shtwo}{\ensuremath{\Sigma_{\rm H_2}}}
\newcommand{\ssfr}{\ensuremath{\Sigma_{\rm SFR}}}
\newcommand{\ha}{\ensuremath{{\rm H}\alpha}}

\newcommand{\sfu}{\ensuremath{\Sigma_{\rm SFR}^{FUV}}}
\newcommand{\sfh}{\ensuremath{\Sigma_{\rm SFR}^{H\alpha}}}
\newcommand{\ms}{\ensuremath{\rm M_{\odot}}}

\title[Star formation in extreme dwarfs]{Small Bites: Star formation recipes in extreme dwarfs}
\author[Roychowdhury et al.]{
Sambit Roychowdhury,$^{1}$\thanks{E-mail: sambit@ncra.tifr.res.in (SR); chengalu@ncra.tifr.res.in (JNC); skai@sao.ru (SSK); begum@astro.wisc.edu (AB); ikar@sao.ru (IDK)} Jayaram N. Chengalur,$^{1\star}$ Serafim S. Kaisin,$^{3\star}$ Ayesha Begum,$^{2\star}$
\newauthor and Igor D. Karachentsev$^{3\star}$\\
       \\ 
       $^{1}$NCRA-TIFR, Post Bag 3, Ganeshkhind, Pune 411 007, India\\
       $^{2}$Dept of Astronomy, University of Wisconsin-Madison, Madison WI 53706-1582\\
       $^{3}$Special Astrophysical Observatory, Russian Academy of Sciences, N. Arkhyz, KChR 369167, Russia}

\begin{document}
\date{}

\pagerange{\pageref{firstpage}--\pageref{lastpage}} \pubyear{}

\maketitle

\label{firstpage}

\begin{abstract}

We study the relationship between the gas column density (\shi) and the 
star formation rate surface density (\ssfr) for a sample of extremely 
small (M$_{\rm B} \sim -13$, $\Delta V_{50} \sim 30$\kms) dwarf 
irregular galaxies. We find a clear stochasticity in the relation 
between the gas column density and star formation. All gas with 
$\shi \gtrsim 10$\ms~pc$^{-2}$ has some ongoing star formation, but the 
fraction of gas with ongoing star formation decreases as the gas column 
density decreases, and falls to about 50\% at $\shi \sim 3$\ms~pc$^{-2}$.
Further, even for the most dense gas, the star formation efficiency 
is at least a factor of $\sim 2$ smaller than typical of star forming 
regions in spirals. We also find that the ratio of \ha\ emission to FUV
emission increases with increasing gas column density. This is unlikely 
to be due to increasing dust extinction because the required dust to gas ratios
are too high. We suggest instead that this correlation arises because
massive (i.e. \ha\ producing) stars are formed preferentially in 
regions with high gas density.  

\end{abstract}


\section{Introduction}
\label{sec:int}

Models of galaxy formation and evolution generally use semi-empirical ``recipes'' to 
follow the process of star formation \citep[e.g.][]{spr05,gov10}. Typically, star 
formation is assumed to set in only above a ``threshold'' gas (column) density
\sgas\ and beyond that to be proportional to a power of \sgas. This is supported by 
observations of nearby star forming galaxies \citep[e.g.][]{s59,ken98}. However, 
most of these observations are of large spiral galaxies, whereas from the hierarchical 
galaxy formation model one would expect that the first formed systems were much 
smaller than the typical $z \sim 0$ spiral. Here we study the relation between gas 
and star formation in nearby, extremely faint ( M$_{\rm B} \sim -13$, 
$\Delta V_{50} \sim 30$ \kms) gas rich dwarfs.

The dwarf galaxies in our sample are dynamically and structurally very different from
the large spiral galaxies for which the widely used star formation recipes have been
derived. Firstly, in our sample galaxies the rotation velocity is not much larger than the
velocity dispersion \citep[e.g.][]{beg03,beg08a}. Further, the gas does not settle 
into a thin disc; the mean observed axial ratio of the {\it gas} discs is $\sim 0.6$ 
\citep{roy10}. Both this as well as the expectation that negative feedback from 
supernovae would play a more important role in small galaxies \citep[e.g.][]{mac99} 
make it likely that the relationship between the gas density and star formation in 
dwarf galaxies is different from that in spirals.

Observationally, there is another major difference between studies of star formation
recipes in dwarf galaxies and spirals. Molecular gas is almost never detected in dwarf
galaxies \citep[e.g.][]{taylor98}, which means that the gas column density has to be
estimated from the HI column density \shi\ alone. On the other hand, in large spirals,
the star formation appears to be governed by the molecular gas density and to be much less 
(if at all) related to the atomic gas\citep[e.g.][]{won02,ler08}. However, in 
\citet{roy09}~(henceforth R09) we showed that for dwarf galaxies, in regions of 
active star formation, the star formation rate \ssfr\ {\it is} correlated to the 
HI column density, albeit with significant scatter.  R09 also found that there 
was no sharp ``threshold'' for star formation, with star formation proceeding 
at all gas column densities, down to the sensitivity limit of the data. 
Similarly, \citet{big10} find that in the HI dominated outskirts of spiral 
galaxies, the SFR and HI are correlated, albeit with a scatter.

In this paper we extend our previous work in two important directions. Firstly we
try to quantify the stochastic nature of the relationship between \ssfr\ and \shi.
Secondly we also study the relationship between the gas column density and the
formation of stars of different masses.

\section{Sample and data analysis}
\label{sec:obs}

Our  sample consists of 23 galaxies drawn from the 
GMRT\footnote{We thank the GMRT staff for having made possible the observations used 
in this paper. The GMRT is run by the National Centre for Radio 
Astrophysics of the Tata Institute of Fundamental Research.}
FIGGS HI 21cm survey 
\citep{beg08} with UV data from 
\emph{GALEX}.\footnote{Some of the data presented in this report were obtained from the 
Multimission Archive at the Space Telescope Science Institute (MAST). 
STScI is operated by the Association of Universities for Research in 
Astronomy, Inc., under NASA contract NAS5-26555. Support for MAST 
for non-HST data is provided by the NASA Office of Space Science via 
grant NAG5-7584 and by other grants and contracts.} See R09 for details. 
For 11 of these 23 galaxies 
\ha~ data from the 6m BTA telescope in Russia is available. The full sample
has median HI mass M$_{HI} \sim 28\times10^6 M_{\odot}$, median blue magnitude M$_{\rm B} \sim -13.2$, and median velocity
width $\Delta V_{50} \sim 32$ \kms. The corresponding values for the \ha\ subsample is M$_{HI} \sim 34\times10^6 M_{\odot}$, M$_{\rm B} \sim -13.5$, $\Delta V_{50} \sim 33$ \kms.
The galaxies with \ha\ observations are listed in Table~\ref{tab:samp}; the 
columns in the table are: 
Column(1)~the galaxy name, 
Column (2)~the absolute blue magnitude (corrected for galactic extinction, 
the internal extinction correction has been assumed to be negligible), 
Column(3)~the distance in Mpc, 
Column(4)~the group membership of the galaxy. All of this data has been 
taken from \citet{beg08}.
Column(5)~the de Vaucouleurs (25 mag/arcsec$^2$) diameter of the optical 
disc.  For dwarf low surface brightness galaxies from the KK lists 
(KK14, KK65, KK144, KKH98), the diameters correspond to the Holmberg 
system (~26.5 mag arcsec$^{-2}$).
Column(6)~the optical axis ratio. Data for columns (5) and (6) have been 
taken from taken from \citet{kar04}.

\begin{table}
\begin{center}
\caption{ The sample}
\label{tab:samp}
\begin{tabular}{|lccccccccccc|}
\hline
Galaxy&M${\rm{_{B}}}$&Dist&Group&a&b/a\\
      &(mag)         &(Mpc)&~&($^\prime$)\\
\hline
\hline 
UGC 685 &$-$14.31&4.5&Field  &1.4$~~$&0.71\\
KK 14   &$-$12.13&7.2&N672   &1.6$^+$&0.37\\
UGC 3755&$-$14.90&6.96&Field &1.7$~~$&0.59\\
KK 65   &$-$14.29 &7.62&Field&0.9$^+$&0.56\\
UGC 4459&$-$13.37&3.56&M81   &1.6$~~$&0.87\\
UGC 6456&$-$14.03&4.3&M81    &1.5$~~$&0.53\\
KK 144  &$-$12.59&6.3&CVn~I  &1.5$^+$&0.33\\
DDO 125 &$-$14.16&2.5&CVn~I  &4.3$~~$&0.56\\
UGC 7605&$-$13.53&4.43&CVn~I &1.1$~~$&0.73\\
DDO 181 &$-$13.03&3.1&CVn~I  &2.3$~~$&0.57\\
KKH 98  &$-$10.78&2.5&Field  &1.1$^+$&0.55\\
\hline
\hline
\end{tabular}
\end{center}
\begin{flushleft}
$^+$: diameters correspond to the Holmberg system (~26.5 mag arcsec$^{-2}$)
\end{flushleft}
\end{table}

\begin{figure*}
\psfig{file=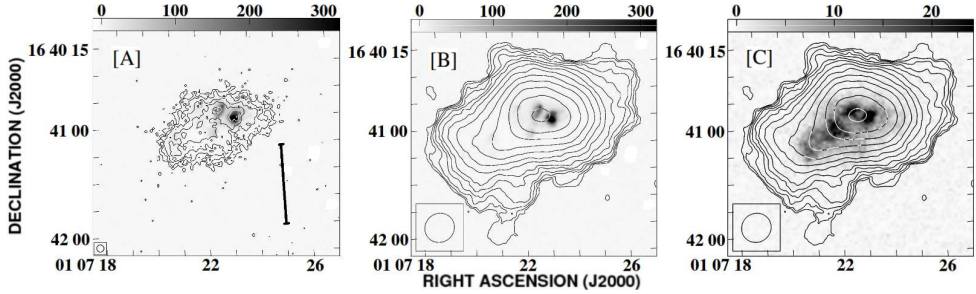,height=1.5truein}
\caption{Overlays of the \ha, UV and HI images for UGC685. 
[A]Greyscales \ha~(in $10^{-18}$ ergs s$^{-1}$ per pixel of area 0.1225 arcsecond squared)
, contours \emph{GALEX} FUV image (from 0.0014 to 0.032 cps per pixel of area 2.25 arcsecond squared, in steps of 2).
[B]Greyscales \ha, contours GMRT HI image (from 17.5 to 1120 Jy/bm$\times$m/s in steps of $\sqrt{2}$).
[C]Greyscales \emph{GALEX} FUV (in $10^{-3}$ cps per pixel of area 2.25 arcsecond squared), contours GMRT HI image. Respective resolutions are, \ha: 1.9\arcs, FUV: 4\arcs, HI: 17\arcs$\times$16\arcs.
The length of the bold line in panel [A] is approximately 1 Kpc.}
\label{fig:ovl}
\end{figure*}

Background corrected \emph{GALEX} FUV band (1350-1750 $\rm{\AA}$) images 
were converted into luminosity units using the calibration information 
provided at the \emph{GALEX} site. Correction for galactic  extinction 
was done using extinction values of \citet{sch98} and using formulae from
\citet{car89} to extrapolate to the FUV band. No correction for internal 
extinction was made, since our sample galaxies are expected to be extremely
dust poor.  The luminosity values thus obtained were converted to star formation
rates
using the calibration given in \citet{ken98a} :
\begin{equation}
\rm {SFR(M_{\odot}~year^{-1})~=~1.4\times10^{-28}~L_{\nu}~(ergs~s^{-1}~Hz^{-1})}
\label{eqn:cal}
\end{equation}
\noindent
In deriving this calibration it is assumed that the stellar distribution
has solar metallicity and a Salpeter IMF, and that the galaxy has had 
continuous star formation over time scales of 10$^8$ years or longer. 
The implications of these assumptions are discussed in 
Section~\ref{sec:res}.

Details of the \ha~ data reduction can be obtained from \citet{kar07} and 
\citet{kai08}. The images were corrected for dust extinction due to our 
own Galaxy in a similar way as was done for the FUV maps. The \ha~ luminosity 
was converted to star formation rates using the calibration given in \citet{ken98a} :
\begin{equation}
\rm {SFR(M_{\odot}~year^{-1})~=~7.9\times10^{-42}~L_{H\alpha}~(ergs~s^{-1})}
\label{eqn:cal2}
\end{equation}
\noindent
The assumptions used to derive this calibration are the same as that used
in deriving the FUV flux -SFR calibration

For data from all the three wavelengths, relevant parameters (\shi~ and \ssfr) 
were calculated over several scales, viz. 
a) an average over the entire star forming disc of the respective galaxy 
(i.e. ``global'' values). The ``star forming disc'' is defined
as that within the radius at which the star formation rate is  1.85$\times$10$^{-4}$ M$_\odot$ yr$^{-1}$ kpc$^{-2}$(as measured
from the FUV flux, with the \emph{GALEX} images smoothed to 400~pc linear resolutions). 
This approximately corresponds to the B~band Holmberg diameter for those
sample galaxies for which the Holmberg diameter has been measured.
b) ``pixel'' values. We use ``pixels'' that Nyquist sample squares 
400~pc or 150~pc in size. For the HI images 400~pc resolution images are 
available for all the galaxies in our sample. Similarly for the FUV data, 
150~pc resolution images are available for all galaxies.

Figure~\ref{fig:ovl} shows \ha~ greyscale images overlayed with FUV and HI 
contours, for a representative galaxy in our sample.

\section {Results and Discussion}
\label{sec:res}

\begin{figure*}

\psfig{file=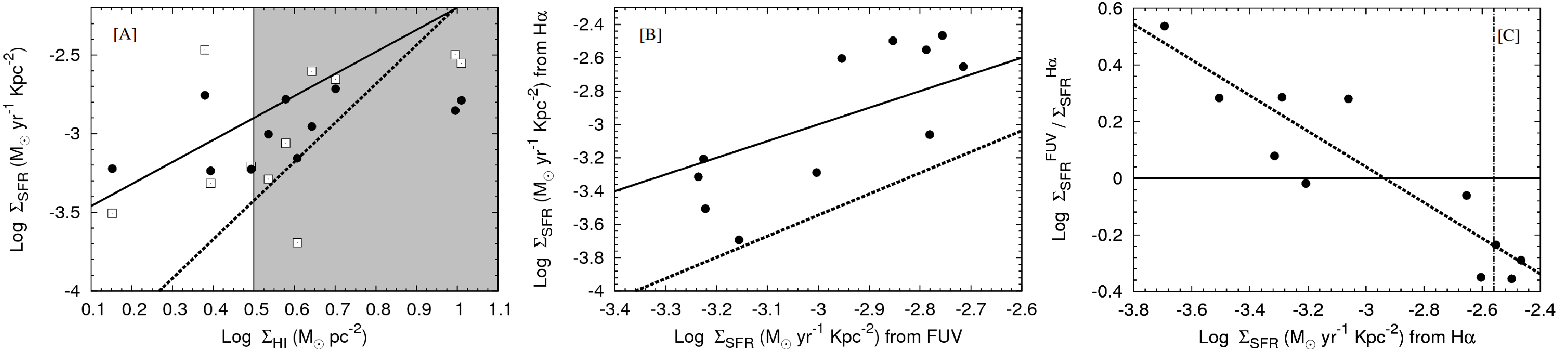,height=1.4truein}
\caption{
[A]\ssfr~ derived from \ha~ (\sfh, empty squares) and FUV 
(\sfu, filled circles) plotted against \shi, assumed to represent \sgas, 
both axes being in log scale. The solid line represents the 
Kennicutt-Schmidt law with a slope of 1.4, and the dashed line represents 
the best fit Schmidt law for spiral galaxies only, both taken from 
\citet{ken98}. The shaded region covers various estimates of the 
``threshold density'' tabulated in \citet{ken89}.
[B]Disc-averaged values of \sfh\ and \sfu. The solid line is the 
1:1 line, and the dashed line represents the relationship 
found by \citet{lee09}.
[C]Ratio of global \sfu/\sfh as a function of \sfh. The dashed line 
is the best fit straight line and has a slope of -0.63. The vertical 
dot-dashed line shows the approximate \ssfr~ value for our sample 
galaxies for which the SFR estimated assuming a Salpeter IMF will 
start deviating from the true SFR according to \citet{pfl07}. See
the text for more details.
}
\label{fig:global}
\end{figure*}

Figure~\ref{fig:global}[A] shows the relationship between the 
disc-averaged \sfh\ (and corresponding \sfu)  and \shi\ for the galaxies 
in our sample with \ha~ data. Note that the galaxies are forming stars even 
though their typical gas density is at or below the 
``threshold density''. Panel~[B] shows how the global \ssfr~ estimates 
obtained using the two different tracers relate. Although the SFR tracers 
do correlate, there is a considerable scatter about the 1:1 line. Note that 
the data agrees better with the original calibration suggested by 
\citet{ken98a} than with the re-calibration suggested by \citet{lee09}, 
though it should be noted that the latter sample is much larger than ours.
In terms of total SFR the values range from $2.79~\times~10^{-4}~{\rm M_{\odot}~yr}^{-1}$ to $1.05~\times~10^{-2}~{\rm M_{\odot}~yr}^{-1}$ with FUV as tracer, 
and from  $2.33~\times~10^{-4}~{\rm M_{\odot}~yr}^{-1}$ to $1.21~\times~10^{-2}~{\rm M_{\odot}~yr}^{-1}$ with \ha~as tracer.
Finally, following \citet{hun10} we show in Panel~[C] the ratio \sfu/\sfh~ 
as a function of \sfh.  There is a clear correlation and
the best fit line has a slope of $-0.63\pm0.09$, (compared to -0.59 
obtained by \citet{hun10}). The SFR calibration we used assumes solar
metallicity, however, as discussed in detail by \citet{hun10}, the fact
that the dwarf galaxies have lower than solar metallicity has 
only a marginal effect on the \sfu/\sfh~ ratio, since both 
calibrations are similarly affected. 

 In what follows we take a look at the relationship between
gas and star formation on small scales, by making ``pixel-by pixel''
comparisons of \shi\ and \ssfr. We first focus on stochasticity
in the star formation and return to the comparison
between \sfh\ and \sfu in Sec.~\ref{ssec:massive}.

\subsection{Stochasticity in Star Formation}
\label{ssec:uv}

    R09 showed that from a comparison of the FUV and HI images,
in star forming regions \sfu\ and \shi~ are related as

\begin{equation}
  \log\sfu = (1.81 \pm 0.05)\log\shi - 4.70 \pm 0.05
\label{eqn:sfu}
\end{equation}

\noindent By comparison with the canonical K-S law

\begin{equation}
  \log\ssfr = (1.4 \pm 0.15)\log\sgas - 3.60\pm0.14
\label{eqn:ks}
\end{equation}.

\noindent and noting that (i)\shi\ is a strict lower limit to the 
total \sgas\, and (ii) for a given FUV flux the inferred SFR
decreases with decreasing metalicity, the robust 
conclusion that one can draw is that the star formation process 
in dwarf galaxies is significantly less efficient than that in 
big galaxies. R09 also showed (see their Fig.~6) that the data
implied stochasticity and were best modelled as a stochastic power
law with a variation of 50\% in the coefficient (as opposed to the slope)
of the power law. \citet{beg06} had also highlighted the stochasticity
in the relation between \ssfr\ and \shi\ in dwarf galaxies. To properly 
characterize the star formation process, one would hence also need 
to know the average fraction of the gas that is participating in 
the star formation process.

\begin{figure}
\begin{center}
\psfig{file=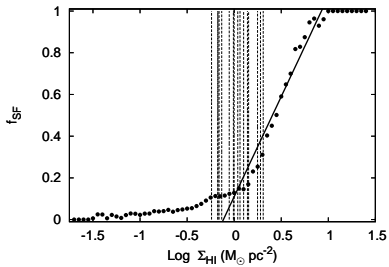,width=1.5truein,angle=270}
\end{center}
\caption{ Plot of the fraction of gas that is ``participating in star
formation'' as a function of \shi. The plot averages over 16 galaxies, the 
dashed  horizontal lines indicate the rms level of the UV images of
the individual galaxies (after being translated from \ssfr\ to \shi\
using Eqn.~\ref{eqn:sfu}). All points to the right of the rightmost dashed line can 
be regarded as giving a reliable fraction of gas that is 
participating in star formation. The solid line is a fit to the
``reliable'' points. See text for more details.}
\label{fig:sfr_frac}
\end{figure}

Figure~\ref{fig:sfr_frac} shows the fraction of pixels which are 
observed to be star forming (i.e. have a star formation rate of at 
least $3\sigma$, where $\sigma$ is the rms  in the UV image, in units 
of the star formation rate).  The plot averages over 16 of the original 
sample of 23 galaxies, 7 galaxies with relatively low GALEX exposure times 
have been excluded. The dashed  vertical lines indicate the rms 
level (after being translated from \ssfr\ to \shi\ using 
Eqn.~\ref{eqn:sfu}). For a given galaxy, if there was no stochasticity in the 
star formation, all points above the rms level (right of the corresponding dashed line) should
have had observable star formation. As such, all points to the right of the 
rightmost dashed line can hence be regarded 
as giving a reliable fraction of gas that is participating in 
star formation. There are several points worth noting, viz.
(1)~all pixels with gas density greater than $\sim 10$\ms/pc$^2$ participate
in star formation. Interestingly, this number is identical to the
threshold density for star formation of $\sim 10^{21}$atoms/cm$^2$ 
proposed by \cite{ski87},
(2)~the fraction of gas which participates in star formation decreases
nearly linearly with decreasing \shi\, (f$_{\rm SF} = 0.96~{\rm log}\shi + 0.1$).
(3)~even for a gas density ${\rm log}\shi \sim -1.0$, two orders of magnitude below
the usually assumed threshold for star formation, at least 5\% of the gas is
observed to be forming stars.
The average \shi\ for the pixels with $\shi > 10$\ms/pc$^2$ is $\sim 17.3$
\ms/pc$^2$, and the average \ssfr\ for these pixels is 
$3.5\times 10^{-3}$\ms/yr/kpc$^{2}$. Thus even for the densest gas in 
dwarf galaxies, the star formation efficiency (i.e. \ssfr/\sgas) is 
hence $\sim 2.0 \time 10^{-10}$yr$^{-1}$, about a factor of two lower 
than the typical value for spiral galaxies \citep{ler08}. 

\subsection{Massive star formation}
\label{ssec:massive}

\begin{figure}
\begin{center}
\psfig{file=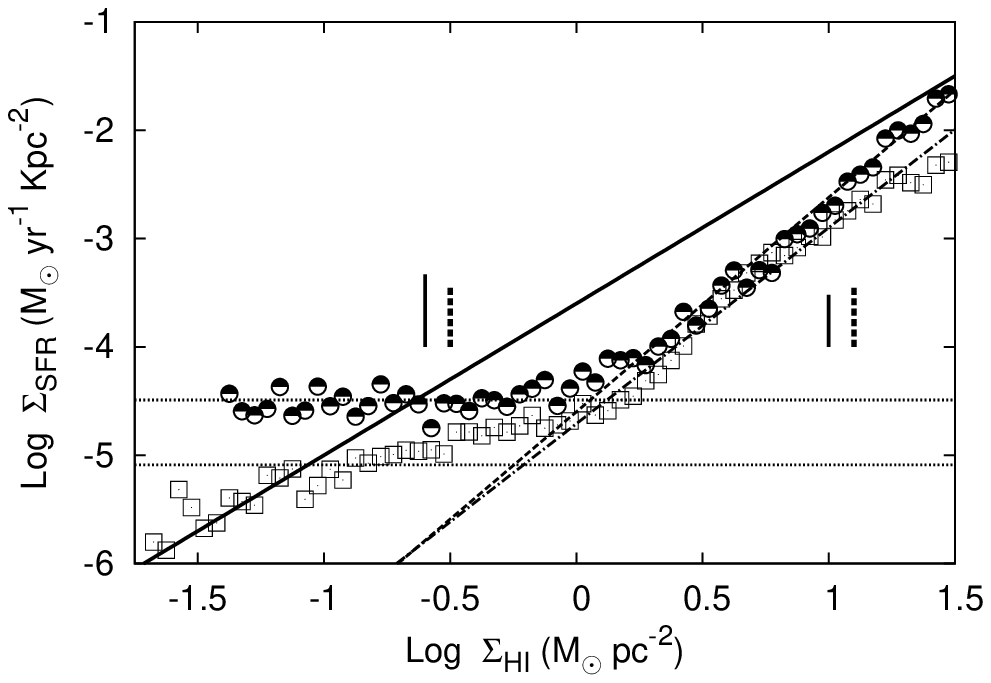,height=1.5truein}
\end{center}
\caption{Plots showing the binned 400pc resolution \sfu~ (hollow squares) 
and \sfh~ (half-filled circles) as a function of \shi.  The \sfu\ data are for the 
23 galaxies listed in R09, where as the \sfh\ data are for the 11~galaxies 
listed in Table~\ref{tab:samp}. The average sensitivity levels for the 
two sets of data are indicated by the respective horizontal lines. The
dashed and the dot-dashed lines show the Schmidt law fit to the 
\ha~ and FUV data respectively. The solid line is the 
canonical K-S law (Eqn.~\ref{eqn:ks}). The typical 1 $\sigma$~scatter 
either above or below the mean in each bin is indicated by the vertical lines,
in (1) the power law and (2) sensitivity limited parts for both tracers.
Dashed lines are for \ha~data, bold lines are for FUV data.
Note that points with negative or zero FUV flux have been dropped from
the plots.
}
\label{fig:sfr_cmp}
\end{figure}

FUV emission is sensitive to the SFR of 
intermediate mass ($M \gtrsim 3$\ms) relatively long lived (lifetime 
$\sim 10^8$yr) stars. \ha\ emission on the other hand traces the 
instantaneous SFR of massive ($M \gtrsim 17$\ms) short lived 
(lifetime $\sim 10^6$yr) stars. For our sample galaxies we show 
in Fig.~\ref{fig:sfr_cmp}  the \ssfr\ as deduced from the FUV 
emission (\sfu), as well as \ha~ emission (\sfh) as a function of 
\shi\ at a resolution of 400pc. (Note that pixels corresponding
to gas not taking part in star formation, i.e. with negative or zero FUV flux, are not included in 
this plot) The best fit power laws to the \ha\ data is given by:

\begin{equation}
\log\sfh = (1.98 \pm 0.04)\log\shi - 4.60 \pm 0.05
\label{eqn:sfh}
\end{equation}

\noindent As can be seen the \sfh\ and \sfu~ points overlap within the scatter
(indicated by the vertical line).
Nonetheless as a comparison of Eqn.~\ref{eqn:sfh} and Eqn.~\ref{eqn:sfu}
shows, there is a significant difference ($\sim 2.7\sigma$, 
where $\sigma$ is the quadrature sum of the individual errors) 
in the slope of the two relationships, with the \sfh\ relation 
being steeper. 

\begin{figure*}
\psfig{file=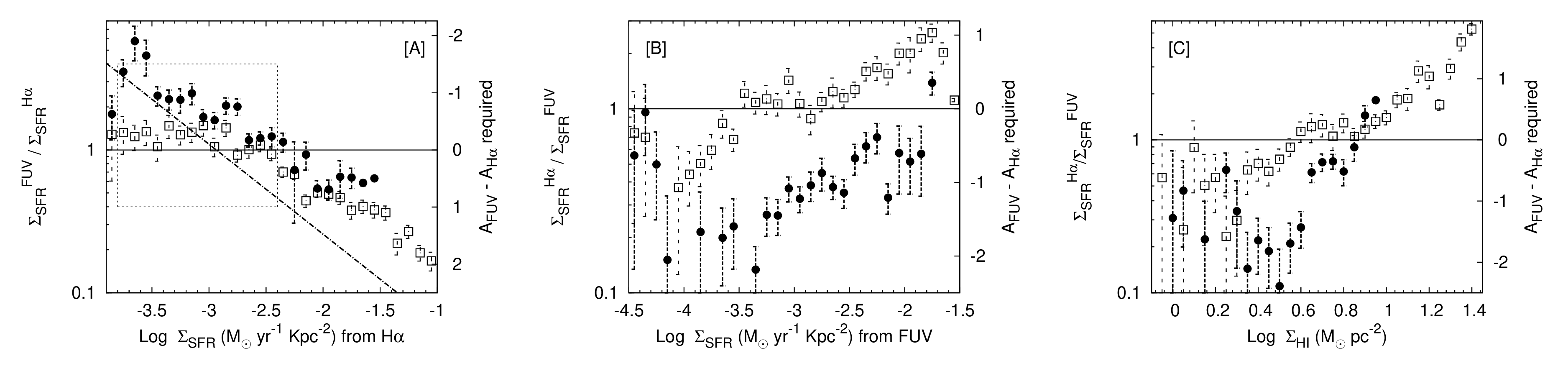,width=1.65truein,angle=270}
\caption{The ratio of \sfh\ and \sfu\ (at scales of 150 pc) plotted as a
function of \sfh\ (Panel[A]), \sfu\ (Panel[B]) and \shi~ (Panel[C]). The
\shi\ is computed from 400pc resolution images. The solid dots are for
for galaxies for which the global $\sfu > \sfh$, and the hollow
square are for the galaxies with global $\sfh > \sfu$.  Means and 
the errors on the mean for each bin are shown. The axes on the right 
show the amount of differential dust extinction required to bring the two 
SFR estimators into agreement. When binning in 
\sfh\ all pixels with \sfh~ more than 3 times the rms value (of the corresponding galaxy) are 
considered (regardless of the value of \sfu), similarly, when binning 
in \sfu\ (and binning in \shi) all pixels with \sfu~ more than 3 times the rms value are 
considered (regardless of the value of \sfh). In Panel[A] the dotted 
rectangular region marks the area covered in the similar global 
plot Figure~\ref{fig:global}[C]. The dashed line in it is the best fit 
straight line from the global plot Figure~\ref{fig:global}[C].}
\label{fig:rfp}
\end{figure*}

Discrepancies between the SFR rates deduced between these two tracers
have been investigated earlier by several authors, including, for e.g.
\cite{meu09,pfl09,lee09}.  A number of explanations for the two rates
to diverge have been suggested, including (i)~Stochastic paucity of high
mass stars at low star formation rates. This would make \sfh\ at low
star formation rates lower than \sfu. For example \citet{lee09} show
that for star formation rates lower than $\sim 10^{-2}$\ms/yr, the
\ha\ emission systematically under predicts the true SFR. (ii)~Non uniform star 
formation rates. For example if the star formation is bursty, then
a few million years after the burst all the OB stars would have died
and the \ha\ emission would once again systematically underestimate the true 
SFR. (iii)~leakage of ionizing photons, either out of the galaxy, or into
a more diffuse region of the ISM, where the resulting \ha\ emission
has too low a surface brightness to be detected (e.g. \citet{mel09}). Once
again, this would result in the \ha\ emission underestimating the
true SFR. (iv)~variations in the IMF.  For example, \citet{meu09} identify
correlations between \sfu/\sfh with global galaxy parameters like
the luminosity, rotational velocity and dynamical mass, and argue
that this implies an IMF that varies with environment. \citet{wei05}
present a model in which the underlying IMF is universal, but 
a dependence of the most massive star formed in a cluster on the 
mass of the cluster leads to the total stellar population having a
steeper IMF than the canonical one. (v)~Dust extinction. Since 
dust extinction is more at the shorter wavelengths, under 
correction for dust would lead to the FUV emission underestimating the true SFR.
Note that in most of the above scenarios  the \ha\ emission would
under predict the true star formation rate. One would expect the \sfh\
to exceed \sfu (as observed for about half of our sample) only if 
(i)~the IMF is more top heavy than assumed, or (ii)~the dust extinction 
has been underestimated. 

   To explore this issue further, we show in Figure~\ref{fig:rfp}, pixel by
pixel correlations of \sfh/\sfu (both at 150 pc resolution) with \ssfr\ and \shi(at 400 pc resolution). In each panel
the hollow squares are for those galaxies for which the global \sfh\ 
is greater than the global \sfu, while the filled circles are for those 
galaxies for which the global \sfh is less than the global
\sfu. From the first panel, one can clearly see that the anti correlation
between \sfu/\sfh and \sfh\ seen on global scales continues even on scales as small
as 150pc. The right axis of the panels is the amount of differential
dust obscuration required to bring the two SFR
estimators into agreement. From Figure~\ref{fig:rfp} one can see that bringing
the two SFR estimators into agreement at the lowest star formation rates
requires the dust obscuration to be more at \ha\ than at FUV, which is
physically implausible. It is more likely that one of the several 
mechanisms discussed above for suppressing the \ha\ flux at low star formation
rates is operative. At high star formation rates, where $\sfh  > \sfu$,
the average N$_{\rm HI}$/A$_{\rm V}$ required to bring the two estimators 
into agreement is 8, i.e. the gas should be about twice as dust rich as the SMC 
(for which N$_{\rm HI}$/A$_{\rm V}$ is 16.3 from \citet{bou85}). If
one assumes that these regions have substantial molecular gas,
and that the galaxies follow the L-Z relation for dwarfs (e.g. \citet{ekta10}),
and that dust is proportional to metalicity, then the required  molecular
gas densities to bring the gas to dust ratio to the same value as the
SMC is $\shtwo \gtrsim 10^2\ms/{\rm pc}^{-2}$, similar to the peak 
densities in the center of spirals, which again seems unlikely. In summary 
it does not appear that dust extinction is the primary cause of the 
disagreement between \sfh\ and \sfu\ at the high \ssfr\ end. 

In terms of direct observables, panel~[B] shows that for the same amount 
of FUV emission, galaxies with lower global \sfh/\sfu are under 
producing \ha\ emission. This could either be because the
galaxies have a fading starburst or because the galaxies are not
producing high mass stars. \citet{lee09,lee09a} find that 
the frequency and amplitude of star bursts in dwarfs make the
former explanation unlikely. However, a more detailed calculation,
and observations of a larger sample would be needed to properly
settle this issue. The most striking feature of the plots however
is in  panel~[C], which shows that galaxies with lower global \sfh/\sfu~ do 
not have gas with column density $\gtrsim 10$\ms/yr. 
The most straight forward interpretation of this is that massive 
star formation is more likely to happen in gas with high column 
densities. Indeed, star formation models have supported such a 
correlation (e.g. \citet{kru10}). While the linear scales that 
the models refer to are much smaller  than those that we are 
dealing with here, such a correlation is likely given that 
high density star forming regions are more likely to occur
in regions where the overall gas density is higher.

\section{Conclusions}
\label{sec:conc}

We find a clear stochasticity between the \shi\ and \ssfr.
All gas with $\shi \gtrsim 10$\ms~pc$^{-2}$ has associated star
formation. While the fraction of star forming gas decreases with
decreasing \shi\ there is no sharp ``threshold'' below which
star formation is completely quenched. We also find that galaxies for 
which globally $\sfu < \sfh$ are marked by not having high HI column density
(i.e. $\shi > 10$\ms~pc$^{-2}$) gas. This is consistent with models
in which formation of high mass stars preferentially happens in regions
with high gas column density.

\bsp

\label{lastpage}

\end{document}